\documentclass[conference]{IEEEtran}
\IEEEoverridecommandlockouts
\usepackage{cite}
\usepackage{amsmath,amssymb,amsfonts}
\usepackage{algorithmic}
\usepackage[ruled,vlined]{algorithm2e}
\usepackage{graphicx}
\usepackage{subfigure}
\usepackage{textcomp}
\usepackage{xcolor}
\usepackage{verbatim}
\usepackage{booktabs}
\usepackage{caption}
\usepackage{makecell}
\def\BibTeX{{\rm B\kern-.05em{\sc i\kern-.025em b}\kern-.08em
    T\kern-.1667em\lower.7ex\hbox{E}\kern-.125emX}}

\begin{document}

\title{   AI-Generated Network Design: A Diffusion Model-based Learning Approach\\
}

\author{
	\IEEEauthorblockN{~Yudong~Huang,~Minrui~Xu,~Xinyuan~Zhang,~Dusit Niyato,~\IEEEmembership{Fellow,~IEEE},\\Zehui~Xiong,~Shuo~Wang,~Tao~Huang,~\IEEEmembership{Senior~Member,~IEEE} \\  }

	\thanks{ Y. Huang and X. Zhang are with the State Key Laboratory of Networking and Switching Technology, BUPT, Beijing, 100876, P.R. China  (e-mail:
		hyduni@bupt.edu.cn, zhangxinyuan0181@bupt.edu.cn)
		
		M. Xu and D. Niyato are with the School of Computer Science
		and Engineering, Nanyang Technological University, Singapore (e-mail:
		minrui001@e.ntu.edu.sg, dniyato@ntu.edu.sg).
		
		Z. Xiong is with Information Systems Technology and Design (ISTD) Pillar, Singapore University of Technology and Design, Singapore (email: zehui\_xiong@sutd.edu.sg).
		
		S. Wang, and T. Huang are with the State Key Laboratory of Networking and Switching Technology, BUPT, Beijing, 100876, P.R. China. They are also with the Purple Mountain Laboratories, Nanjing, 211111, P.R. China (e-mail:
		shuowang@bupt.edu.cn, htao@bupt.edu.cn)

}}

\maketitle

\begin{abstract}
The future networks pose intense demands for intelligent and customized designs to cope with the surging network scale, dynamically time-varying environments, diverse user requirements, and complicated manual configuration. However, traditional rule-based solutions heavily rely on human efforts and expertise, while data-driven intelligent algorithms still lack interpretability and generalization. In this paper, we propose the AIGN (AI-Generated Network), a novel intention-driven paradigm for network design, which allows operators to quickly generate a variety of customized network solutions and achieve expert-free problem optimization. Driven by the diffusion model-based learning approach, AIGN has great potential to learn the reward-maximizing trajectories, automatically satisfy multiple constraints, adapt to different objectives and scenarios, or even intelligently create novel designs and mechanisms unseen in existing network environments. Finally, we conduct a use case to demonstrate that AIGN can effectively guide the design of transmit power allocation in digital twin-based access networks.

\end{abstract}


\section{Introduction}

The future networks have attracted much attention due to the unprecedented capabilities of bearing the Metaverse, satellite-terrestrial integrated networks, connected vehicles, and other emerging applications\cite{6GWS}. However, with the rapid growth in network scale and user demands, recent years have witnessed dramatic increases in the complexity and dynamics of network designs. For instance,  the cross-domain network planning requires a lot of expert knowledge and experience, and layered protocol designs heavily rely on human efforts. Moreover, in large-scale time-varying scenarios, it is intractable to construct a complete system model with appropriate constraints\cite{NI6G}. Faced with enormous operation and maintenance costs, the future networks pose intense demands for intelligent, automated, and customized network designs.

Recently,  artificial intelligence-generated content (AIGC) provides an exciting perspective for intention-driven network design. Based on generative AI techniques and applications, such as diffusion models\cite{ddpm} and ChatGPT\cite{openai2023gpt4},  AIGC exhibits outstanding inference capabilities in text processing, code writing, and text-to-image, where humans interact with intelligent machines through natural language to efficiently complete design tasks.  In particular, diffusion models have been successfully leveraged to optimize the decision-making in reinforcement learning (RL), where we can encode the design intention into the conditioning information and explicitly guide the generation of  realistic trajectories (i.e., state-action-reward sequences that represents the decision-making process) with flexible combination of multiple constraints. Although datasets collected from network systems are not normalized data like image pixels, many network optimization problems are viable to model as the decision-making process of RL. 

 On the other hand, previous studies mainly focused on solving network optimization problems by using intelligent algorithms, such as deep reinforcement learning (DRL)\cite{neplan} and federated learning, while the lack of interpretability and generalization makes intelligence-based networking systems prohibitive to deploy in practice. Digital twins are expected to narrow this gap by verifying the algorithms in virtual worlds that map from physical infrastructure. With the advances of generative AI,  RL, and digital twins, we envision a paradigm shift from rule-based manual design towards intention-driven intelligent design for  future networks. 

In this paper, we propose the AIGN (AI-Generated Network), a novel network design paradigm that can automatically generate customized solutions to adapt to dynamically time-varying environments. Unlike traditional optimization methods, AIGN does not require building a complete network system model and knowing all constraints prior.  Moreover, by merging the generative ability of diffusion model with the RL, AIGN is promising to achieve intelligent decision-making in many network design scenarios, such as satellite-terrestrial network planning and mobile ubiquitous computing.

The main contributions of this article are as follows.
\begin{itemize}
\item We present the novel AIGN framework and analyze its appealing properties, including scalability for large-scale optimization, flexibility for multi-constraint combination, and interpretability for intent-driven configuration.

\item To enable the AIGN, we propose a diffusion model-based learning approach, where the critical  techniques for generating novel network designs are detailed.

\item We conduct a proof-of-concept simulation in digital twin-based access networks. The results show that AIGN can effectively guide the design of transmit power allocation by learning the reward-maximizing trajectories.

\item We discuss the challenges and potential research directions brought by the AIGN, such as high-quality datasets, performance metrics, and interaction platforms.

\end{itemize}

The rest of the article is organized as follows. We commence with the AIGC and network design issues, clarify our motivation, and give the design principles. Next, we present the AIGN framework. Then the key functional components of diffusion model-based learning approach are detailed. Following that, we give the case study and analyze the challenges. Finally, we draw the main conclusions. 
 
\begin{figure*}[]
	\centering
	\includegraphics[width=6.8in]{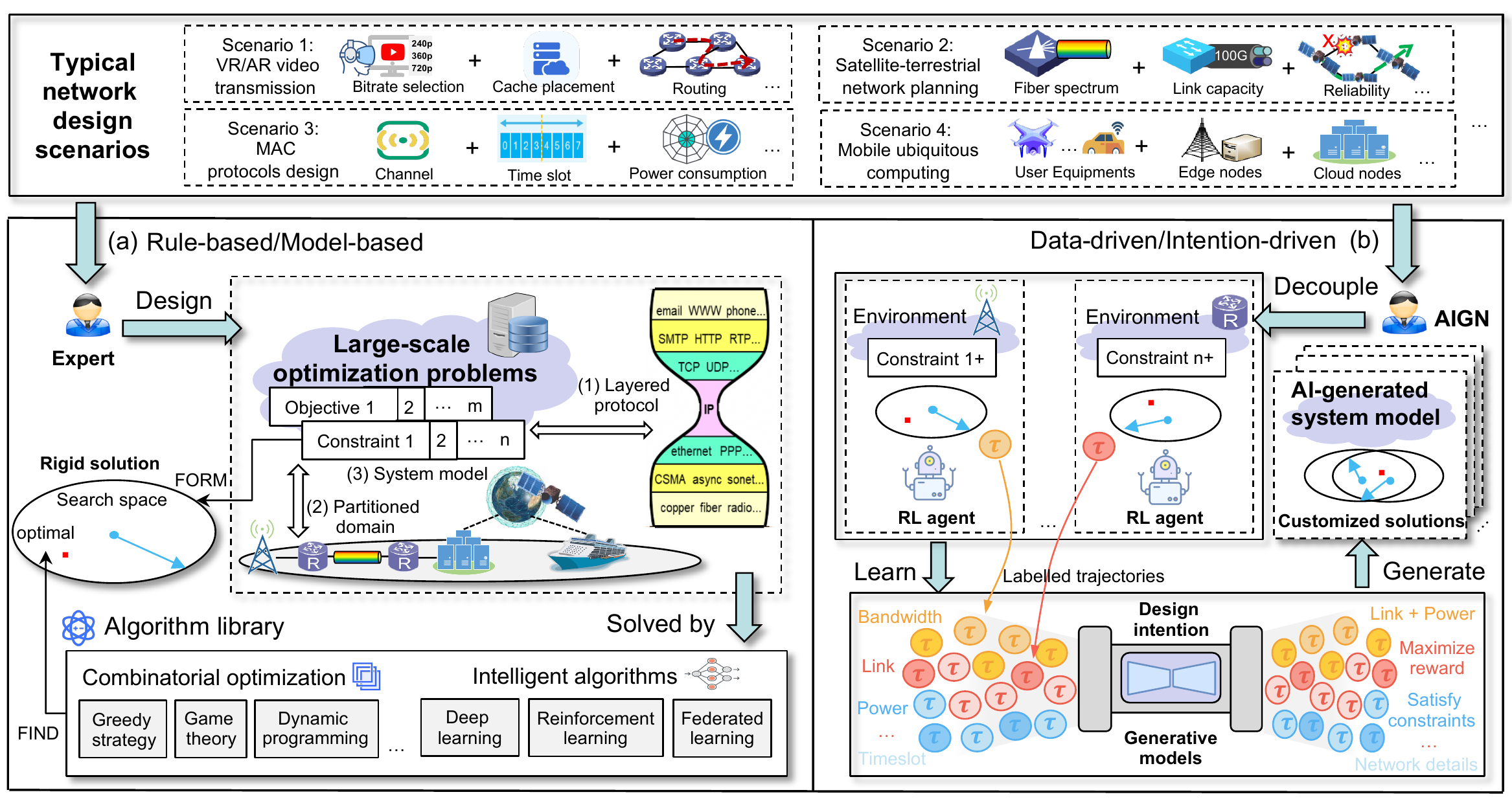}
	\caption{Comparing traditional network expert design with AIGN. (a) Rule-based methods require manually building the system model, and pursue the optimal results that heavily depend on expert knowledge. (b) AIGN prioritizes the decision-making process with high rewards, and memorizes these trajectories as experiences. This allows AIGN to automatically generate various customized designs that are aligned with the intention.}
	\label{fig:AIGN_motivation}
\end{figure*}

\section{AI-Generated Network}

This section overviews the generative AI techniques, and their applications in the network design, emphasizing the motivation and design principles.

\subsection{AI-Generated Content}

AIGC refers to the use of generative AI techniques (e.g., Autoregressive Models, Variational Auto-Encoders, Generative Adversarial Networks, Flow-based Models, and diffusion models\cite{ddpm}) to automatically produce high-quality  customized content. Recently, diffusion models have shown groundbreaking success in the task of text-to-image and decision-making. Thanks to the great potential in creating new patterns, diffusion models have also been applied to generate the molecular conformer and protein structure,  which unveils a new revolution in exploiting the generative AI for science. This paper focuses on the RL-based decision-making problems in network fields. Since the diffusion models have been proven to combine with RL\cite{is_conditional}, we adopt the diffusion models as the main representative of generative techniques, while other generative models may also be utilized.

The technique of generative AI lies in that: Given observed samples $x$ from a distribution of interest (e.g., an image), the generative AI  learns to model its true data distribution $p(x)$. Once learned, it can generate new samples from the approximate model at will. Then,  by explicitly controlling the data generated through conditioning information $y$ (e.g., a text),  we can learn the conditional distribution $p(x\mid y)$ (e.g., a new image in response to text prompts)\cite{luo_diff}.

\subsection{Network Design Issues}
 In the past decades, network protocols have laid the foundation of the Internet, while optimization algorithms and configuration parameters greatly affect network performance.  As shown in Fig. \ref{fig:AIGN_motivation}(a), the current design approach for large-scale networks mainly includes three aspects: (1) The layered protocol is used to simplify the design. Each layer provides a set of communication/networking functions. (2) The network domain is adopted by partitioning networks into regions to improve scalability. (3) Human experts empirically design large-scale optimization problems by specifying the  system models and objective functions with multiple constraints, especially across layers and domains. Based on the rule  and model, the experts can form a search space for the problem, and then use combinatorial optimization (such as greedy iterative strategy) or intelligent algorithms to find the optimal solution (e.g., maximize throughput, minimize power consumption, and reduce hardware device and its deployment expense). 

Nevertheless, future networks not only require full-coverage connections in the air, space, ground, and sea, but also need to provide real-time and high-bandwidth services for ubiquitously connected users. As the network size and number of users proliferate, it is difficult to manually formulate the network problem from a global perspective and directly construct a complete system model with all constraints. Moreover, the traditional rule-based optimal solutions are prone to suffer from dynamic time-varying environments,  where customized solutions that quickly adapt to a variety of realistic situations are more essential than a rigid optimal solution. To be specific, we present four typical scenarios to analyze large-scale network design issues.

\textbf{Use case 1: VR/AR video transmission.} As virtual reality (VR)/augmented reality (AR) and high-definition live stream are applied to education, medical care, entertainment, and business, video traffic occupies most of the current Internet bandwidth. To improve the Quality of Experience (QoE) metrics, content providers struggle with constructing high-quality video transmission networks by systematically considering bitrate selection, cache placement, routing, and other optimization problems. However, peak demands (e.g., concerts and sport events) easily cause network congestion and video interruption, challenging the flexibility and scalability of manual-configured transmission systems, further leading to significant revenue losses for content providers.

\textbf{Use case 2: Satellite-terrestrial network planning.} The satellite-terrestrial backbone network interconnects hundreds of regions and serves hundreds of terabits of traffic at all the time. With the rapidly growing bandwidth demand of emerging applications, such as big data and Metaverse, the operators spend billions of dollars every year to plan and upgrade backbone networks. Given traffic demand forecast and requirements (e.g., failure redundancy for reliability), network planning refers to continuously making cross-layer decisions, including both the optical layer (e.g., fiber path to build, fibers to turn on, spectrum to use) and the IP layer (e.g., link capacity to add and routers to procure), to minimize the network infrastructure expenditure. Moreover, satellite clusters have dynamic topologies and unstable time-varying orbital connections, which further exacerbate the challenges of joint satellite-terrestrial optimization.

\textbf{Use case 3: MAC protocols design.} The wireless medium access control (MAC) protocols are mainly responsible for the channel access of user equipments.  The  protocol design involves complex interaction processes and configuration parameters, which must be tailed to specific purposes and scenarios,  such as improving throughput, reducing power consumption, and guaranteeing fairness. Especially under the heterogeneous network (HetNet) deployment scenarios, each wireless access network, e.g.,  5G-NR,  Wi-Fi,  Bluetooth, Zigbee, and even satellite access network, has its own protocols and properties, such as capacity, delay, coverage, security, access technology, power consumption, and cost. The HetNet poses great challenges for efficiently managing and coordinating multiple types of wireless access technologies. Considering the even more complex and diversified settings in future networks, AI-generated protocol design is essential to relieving humans from these burdensome undertakings\cite{6GAI}.

\textbf{Use case 4: Mobile ubiquitous computing.}  Computing resource allocation plays a particularly important role in ubiquitous connection scenarios, such as Industrial Internet of Things, vehicles, drones, and satellites. Due to the limited on-chip computing resources of local devices, computing-intensive tasks usually need to be offloaded to edge nodes or cloud nodes. The offloading decisions are expected to be globally optimal under complicated system models that include computing, communication, power, mobility, distance, and delay constraints. As concepts such as fog computing, device-to-device computing, in-network computing, and serverless computing continue to be proposed, it is quite intractable for network experts to give  a complete system model or a feasible solution in practice.

\begin{figure*}[]
	\centering
	\includegraphics[width=7in]{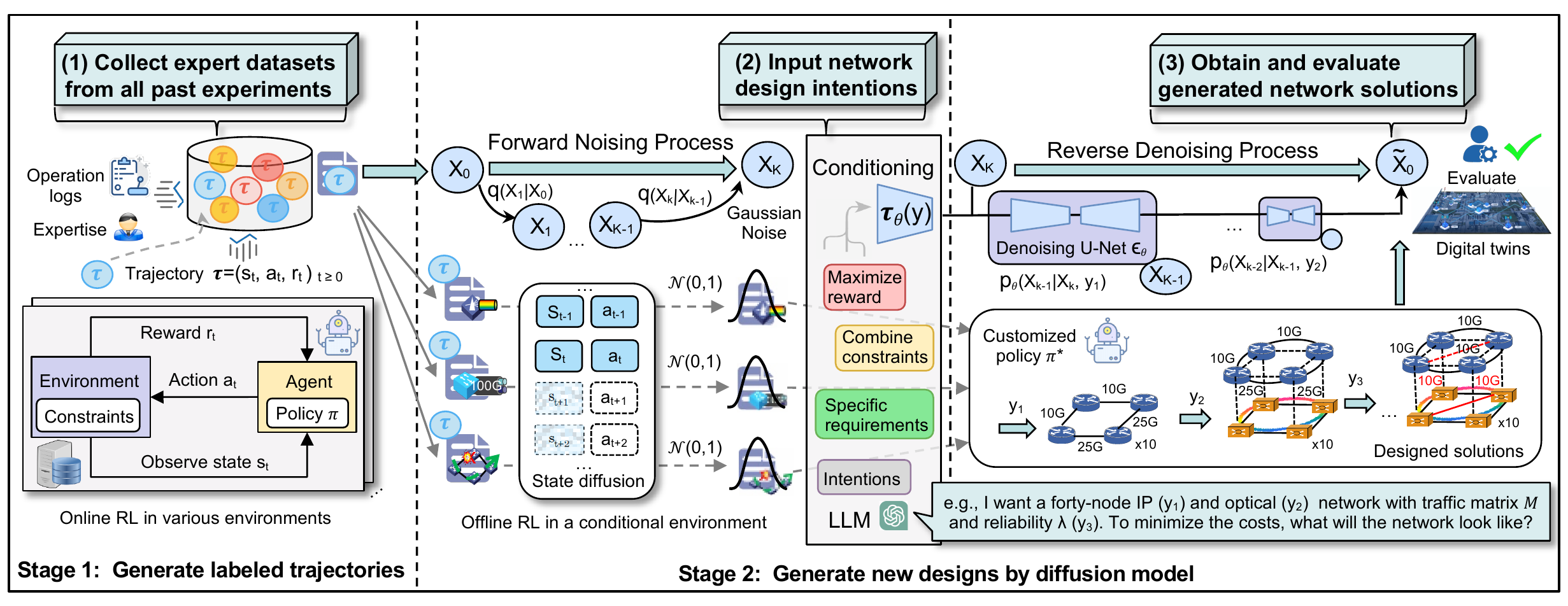}
	\caption{AIGN framework with the diffusion model-based learning approach. After the operator inputs the intentions (e.g., the network capacity planning), AIGN samples from high-reward trajectories and infers the decision-making process from the initial state to the target solution with customized policy $\pi^{*}$.}
	\label{fig:AIGN_design}
\end{figure*}

\subsection{Motivation}

AI techniques profoundly change the  way of design and deployment for future networks. On the one hand, recent studies on AI-assisted network design leverage various Neural Network structures, such as Deep Neural Network, Recurrent Neural Network, and Graph Neural Network, to learn the nonlinear mapping between problem parameters and optimal solutions. Especially,  DRL has been widely used in various communication and network scenarios due to the advantages of label-free training and long-term optimization. On the other hand, there are three significant limitations. First, it still requires expert knowledge to combine multiple network problems to a large-scale optimization problem  and empirically tune numerous learning parameters. Second, the underlying  Neural Networks remain blackboxes where the performance stays uninterpretable for network operators. The lack of interpretability \cite{METIS} makes intelligence-based networking systems prohibitive to deploy in practice. Third,  a model well trained in one environment may perform much worse than traditional rule-based schemes in another new environment. The lack of generalization means endless model design and parameter tuning, which greatly limits the development of intelligence in the network.

To deal with this dilemma,  this paper targets to explore AI-generated methods to assist network design tasks.  Specifically, we propose a new paradigm for the network design, named AIGN. As shown in Fig. \ref{fig:AIGN_motivation}(b), AIGN is a generative network expert system that allows operators to specify customized properties of the network solution and automatically create network design details without requiring the exhaustive system model and constraints. Driven by the holistic data (e.g., channel state, traffic matrix, operation logs, and expertise) and design intentions (e.g., constraints and objectives), AIGN is able to learn new problems from basic optimization problems, automatically combine multiple network subsystems, or even create novel designs and mechanisms unseen in existing network environments, which enhances the interpretability and generalization capabilities of intelligent networks.

\subsection{Design Principles}

Generally, large-scale optimization problems consist of multiple subproblems. Specific subproblems with simple constraints are usually well-studied and RL is viable to generate labeled trajectory datasets for particular network environments.  The authors in \cite{METIS} decompose networking systems into local systems and global systems, and use the decision tree method to interpret a single RL agent's decision-making process in the local system. In \cite{GNET}, the curriculum learning is introduced to improve the generalization of RL. They use rule-based algorithms to generate training curricula. By iteratively training the RL agent in environments where it performs worse than rule-based baselines, the RL agent achieves asymptotic performance in multiple environments (problems). In \cite{WMAC}, the wireless MAC protocol design is decoupled into a set of building blocks, each of which is trained with an RL agent. Multiple agents share their information with a global agent for clustered decision-making. Since the future networks are composed of a large number of heterogeneous and decentralized network entities, multi-agent RL allows each network entity to learn its optimal policy by observing not only the environments but also the policies of other entities, which can significantly improve the learning efficiency. \textit{Thus, our first observation is that large-scale network optimization problems can be transformed into a data-driven paradigm of multi-agent RL\cite{mrl_fe}.}

Then,  the critical challenge is to enable RL agents to intelligently collaborate and learn from each other, so that they can work together to generate new network designs. One of the most powerful solutions is generative AI.  The authors in \cite{NETSHARE} prove the feasibility of using Generative Adversarial Networks (GANs) to automatically generate synthetic packet header traces for networking tasks (e.g., telemetry and anomaly detection). However, these generated traces are purely packet records (such as IP address, port number, and the number of bytes) rather than collected from state-action-reward sets of RL environments; thus they cannot be directly utilized for decision-based problem optimization.  Recently, a Decision Diffuser\cite{is_conditional} framework based on  the diffusion model has been proposed  in the robotic field. Given several RL trajectories collected from simple constraint environments, the Decision Diffuser generates not only typical  behaviors satisfying individual constraints but also new behaviors that maximize rewards by flexibly combining multiple constraints at test time. \textit{Thus, our second observation is that by introducing the inference ability of generative models to the collaboration among multi-agent RL, it is expected to achieve expert-free problem optimization and automatic generation of new network designs.}

\section{AIGN Design}

Motivated by the above issues and trends, we design the AIGN  framework with the diffusion model-based learning approach. Several appealing properties of AIGN are derived, and the key technical components are detailed.

\subsection{AIGN Framework}
As shown in Fig. \ref{fig:AIGN_design}, the workflow of AIGN is composed of the following steps. (1) Collect expert datasets from all past online RL experiments. Data collected from network systems is not normalized data like image pixels, which makes it extremely intractable to unify large-scale network datasets. Fortunately, many network optimization problems can be modeled as the decision-making process of RL, where the trajectories are stored and utilized in a standardized manner. In addition, operation logs and expertise can be encoded into trajectory labels to further improve the quality of datasets. (2) The operator inputs network design intentions (e.g., I want a forty-node IP and optical topology with traffic matrix $M$ and reliability $\lambda $. To minimize the costs, what will the network look like?). The Large Language Model (LLM), e.g., ChatGPT, can be utilized to execute the intention compilation and generate explanations. The intentions are abstracted into optimization objectives, constraints, and specific requirements. Then, the conditional datasets corresponding to intentions are sampled to train an offline RL. With the ability of in-context learning, AIGN infers the design details step by step through dialogue with operators. (3) Finally, the operator obtains a variety of customized network solutions by just clicking a button. The solutions are evaluated in the digital twin-based virtual counterpart to reduce the cost and risk of deployment. As the generative network models continuously evolve,  AIGN has several appealing properties as follows.

\textbf{Scalability for large-scale optimization.} The future networks must become highly scalable  to support a large number of heterogeneous networking scenarios. In practice, a single local network with hundreds of nodes and links may be translated into an ILP (Integer Linear Programming) problem with millions of variables and constraints\cite{neplan}. To achieve joint optimization of heterogeneous networks, even highly-skilled experts require much effort to find an actionable solution. By continuously learning from various network subsystems, AIGN  intelligently synthesizes large-scale optimization problems,  generalizes well to new environments, and  pursues long-term reward maximization, while traditional hand-tuned heuristics are prone to local optima.

\textbf{Flexibility for multi-constraint combination.} In the process of network design, we concern not only about the solution outcomes, but also about the impact of each part on performance and costs. For instance, the wireless MAC protocol design may require  removing the retransmission mechanism to reduce the power consumption, or adding an adaptive sending function to improve the average channel throughput. AIGN provides the service to simultaneously compare alternatives  that are optimized for multiple objectives and constraints, avoiding the tedious trial-and-error modeling process. Consequently, AIGN will greatly accelerate the innovation of network protocols and mechanisms.

\textbf{Interpretability for intention-driven configuration.} After the network has been deployed, operators are often overwhelmed with failures and alarm events. Due to the lack of interpretable intelligence, the existing network management heavily  relies on human analysis and manual configuration. By synchronously sensing the physical network environments with digital twins and learning from historical operating experience, AIGN may intelligently give appropriate operational suggestions, such as increasing link capacity or migrating the path of heavy-loaded applications when network congestion occurs. This transparency of AIGN allows network operators to understand how the system works and make informed adjustments with fewer potential errors. Further, AIGN is expected to realize the intention-driven networks  and automatic network management.

\subsection{Diffusion Model-based Learning Approach}
We make a concrete step towards the AIGN by proposing a diffusion model-based learning approach. The approach is divided into two stages. The first stage is to explore various environments and generate labeled trajectories by online RL agents, and the second stage is to generate new designs by the offline RL with the diffusion model. Note that AIGN is not limited to the diffusion model, and other generative models could also be applied to the AIGN. Next, we present the key technical details and analyze the theory behind AIGN.

\textbf{  Stage 1: Generate labeled trajectories by RL agents.} 
Many communication and network  problems\cite{a_rl_s} can be formulated as a Markov Decision Process (MDP), where an online RL agent is trained to make decisions based on trial-and-error interactions with an environment. At each step, the agent observes the current state $s_{t}$ of the environment and selects an action $a_{t}$ to take, which may lead to a new state $s_{t+1}$ and a corresponding reward $r_{t}$. The goal of the agent is to learn a policy $\pi$ that maximizes the expected cumulative reward over a sequence of state-action-reward trajectory $\tau$. By encoding operation logs and expertise (e.g., returns, specific constraints, and optimization objectives) into trajectories, we can obtain the labeled trajectories that are analogous to well-marked standard image datasets, where generative AI techniques have already found success. 

\textbf{ Stage 2:  Generate new designs by diffusion model.} 
The diffusion model is based on a Markov chain that simulates a forward noising process $q(x_{k+1}|x_{k})$, which gradually adds noise to the orginal data $x_{0}$ until the data  $x_{k}$ becomes a Gaussian distribution $x_{k}\sim \mathcal{N}(0,I)$. Then, by sampling the data  $x_{k-1}$  with the conditioning information $y$, the  reverse denoising process $p_{\theta }(x_{k-1}|x_{k}, y)$ iteratively removes the noise and finally learns the new conditional data  $\tilde{x_{0}}$\cite{ddpm}, i.e., the target data that satisfies user requirements. The temporal U-Net\cite{is_conditional}, a neural network consisting of repeated convolutional residual blocks, can be trained to predict the perturbed noise $\epsilon_{\theta}$ that needs to be removed at each denoising step.

Treating labeled trajectories $\tau$ in online RL as the original data distributions, i.e., $x_{0}({\tau})$, the offline RL with the diffusion model has unprecedented abilities to handle complex high-dimensional state spaces and learn new customized policy $\pi^{*}$ that closely match the expected network solutions. For instance, we can explicitly define the conditioning variables $\left \{ y^{i}(\tau) \right \}_{i=1}^{n}$ as a one-hot vector that represents the returns or satisfied constraints, where $i$ is the indice and $n$ is the number of variables. As shown in Fig. \ref{fig:AIGN_design}, the high-reward trajectories under multiple combinations of IP-layer, optical-layer, and reliability constraints are sampled as experiences. After conditioning on the IP ($y_{1}$), optical ($y_{2}$), and reliability ($y_{3}$), the offline RL with the diffusion model infers the decision-making process from the initial state to the target solution that satisfies multiple constraints at the same time.

\section{Case Study}
In order to evaluate the performance of proposed diffusion model-based learning approach, we create a proof-of-concept prototype under the wireless access scenarios. Simulation results exhibit that AIGN can guide the power allocation. 

\begin{figure}[t]
	\centering
	\includegraphics[width=3in]{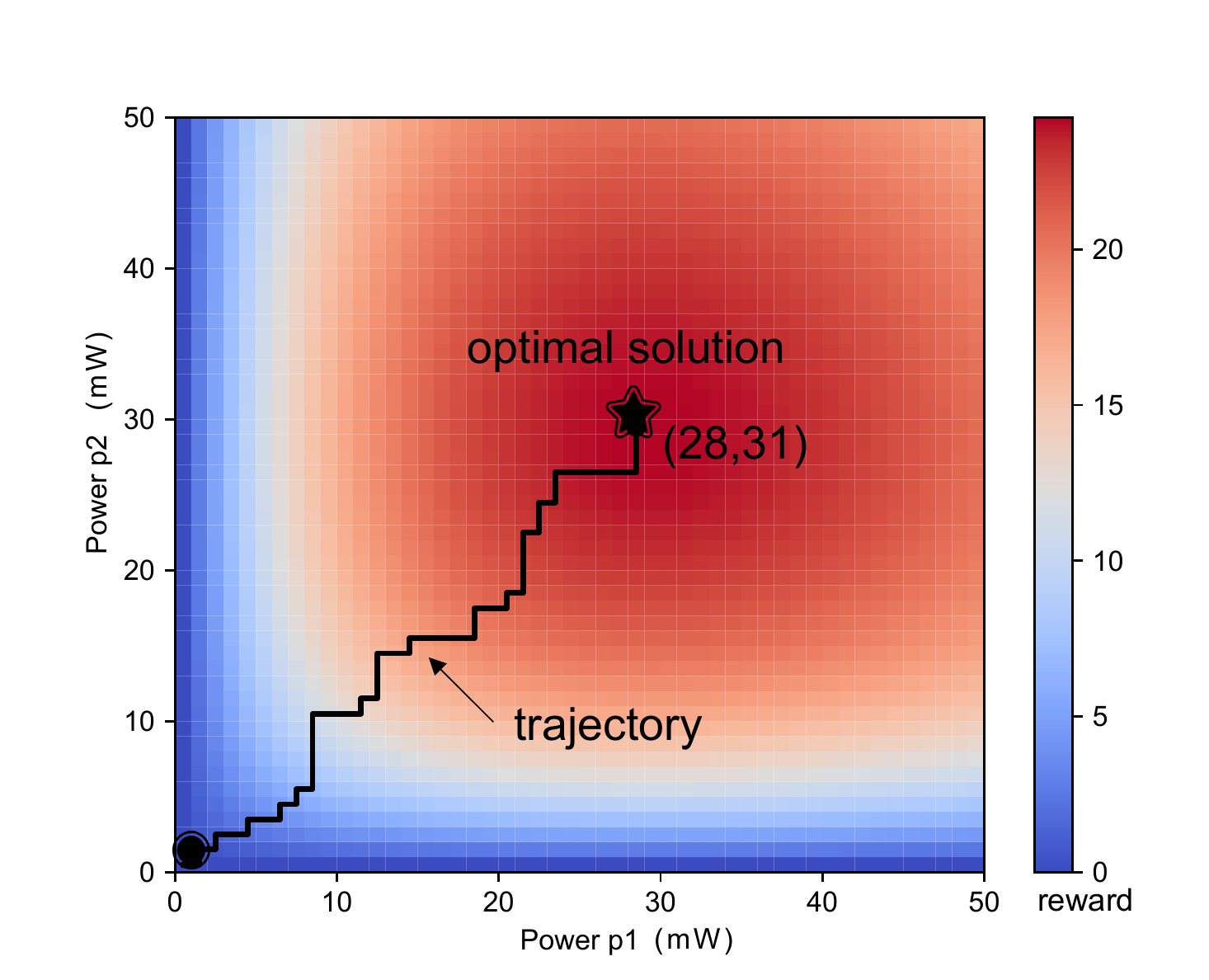}
	\caption{Simulation results for accessing two users. The initial power value is set to (0, 0), and the proposed approach quickly generates the optimal power allocation scheme of (28, 31).}
	\label{fig:2_user}
\end{figure}

\begin{table}[t]
	\centering
	\captionsetup{justification=centering}
	\caption{Multi-user access solutions. The reward values are calculated based on the sum-data utility function.}
	\label{tab:rewards}
	\begin{tabular}{cccc}
		\hline
		Best reward values & 20 Users & 50 Users & 80 Users \\
		\hline
		Soft actor-critic & 254.04 & 488.90& 632.89 \\
		Batch-constrained deep Q-learning & 254.16 & 489.08 & 632.59 \\
		Diffusion model-based learning & 254.15 & 489.02 & 633.02 \\
		\hline
	\end{tabular}
\end{table}

\textbf{Simulation Setup:} The uplink transmission power allocation for sum-data utility maximization with interference in multi-access base stations is a classic NP-hard problem in wireless networks.  We build a digital twin simulation platform to design the optimal discrete power selection scheme ($p_{i,j}$) under time-varying environments with different channel gains $h_{i,j}$, noise power $\sigma^{2}$, the indices of base stations $j$, and the indices of users $i$. The optimization objective is set to maximize the marginal utility, which is a logarithmic revenue  of rate $R_{i,j}$ subtracting the linear power cost $C_{i,j}$. The  rate $R_{i,j}$ is computed according to the signal-to-interference-plus-noise ratio (SINR) and Shannon formula. The  power cost $C_{i,j}$ is proportional to the power $p_{i,j}$.

In stage 1, we train online RL agents to explore various environments with the soft actor-critic (SAC) algorithm. The SAC is a variant of the Actor-Critic algorithm that uses the maximum entropy framework to encourage exploration during training, where the actor learns a stochastic policy that maximizes the expected sum of rewards over time and the critic estimates the Q-values of state-action pairs. These training trajectories are stored in the experience buffer and labeled with information such as returns, constraints, and environmental parameters. In stage 2, we feed about one hundred thousand labeled trajectories to the diffusion model and train the  reverse denoising process of U-Net\cite{is_conditional} with maximum-likelihood estimation.  

\textbf{Simulation Results:} To facilitate understanding and visualization, but without loss of generality, we firstly set  the number of base stations to 1 and the number of users to 2. As shown in Fig. \ref{fig:2_user}, after just inputting the channel gain and noise parameters, the diffusion model quickly generates the reward-maximizing trajectory and finds the best power allocation scheme $(28, 31)$, which proves that AIGN can achieve expert-free problem optimization. Then,  by further conditioning on the user numbers of 20, 50, and 80, AIGN samples from the corresponding conditional datasets and give the transmit power allocation solutions that relate to the  required user numbers. As shown in Table \ref{tab:rewards}, the scheme rewards of the diffusion model-based learning approach are 254.15, 489.02, and 633.02, which are very close to the online SAC and offline batch-constrained deep Q-learning (BCQ) algorithms. As shown in Table \ref{tab:compare}\cite{is_conditional}\cite{under_orl}, both the proposed approach and offline BCQ algorithms learn directly from the datasets, which have faster convergence speed than online RL. Offline BCQ is less time-consuming but the performance highly depends on the quality of data distribution. If the dataset is biased or incomplete, the agent will learn suboptimal policies or even fail to learn. Our approach also requires large and diverse datasets to avoid overfitting, and the conditioning generation ability of the diffusion model helps the agent to overcome the dataset bias and adapt to conditional environments.  Due to the limit of space, we leave more simulations based on AIGN as future work.

\begin{table}[t]
	\centering
	\captionsetup{justification=centering}
	\caption{Comparison of SAC, BCQ and the proposed diffusion model-based learning approach.}
	\label{tab:compare}
	\begin{tabular}{cccc}
		\hline
		Methods & SAC & BCQ & Proposed \\
		\hline
		Category& Online RL & Offline RL& Offline RL \\
		\hline
		Features &{\makecell[c]{Interact with\\ environments}}   & \makecell[c]{Learn from \\  datasets} &  \makecell[c]{Learn from \\  datasets} \\
		\hline
		Convergence & Slow and unstable  & Fast & Fast \\
		\hline
		Pros & {\makecell[c]{ Suitable for \\changing\\ environments}} &  {\makecell[c]{Less time-\\consuming}}&   {\makecell[c]{  Adapt to \\conditional\\ environments }}  \\
		\hline
		Cons & {\makecell[c]{ Time-\\consuming}} & {\makecell[c]{  Dataset bias \\and overfitting}}&   {\makecell[c]{  Require more \\tuning }}  \\
		
		\hline
	\end{tabular}
\end{table}

\section{Challenges and Future Prospects}
AIGN paves the way for intelligent and automated network design, while leaving some challenges to be discussed. In this section, we analyze the challenges brought by the AIGN paradigm and highlight the potential research directions.

\textbf{Realistic sources for high-quality datasets:} Realistic data sources collected from physical infrastructure (e.g., ground-truth channel states and traffic matrix)  are the best datasets to guarantee the fidelity and accuracy of learning trajectories. Besides, synthetic datasets produced by simulators, expertise, or other mixed methods are promising alternatives to  overcome the data access barrier and mitigate the privacy concerns. The dataset characteristics, which heavily influence the performance of AIGN, still remain to be studied. To evaluate the quality of different datasets, the Trajectory Quality (TQ) defined by the average dataset return and the State-Action Coverage (SACo) calculated by the number of unique state-action have been proposed\cite{under_orl}. Moreover, reinforcement learning from human feedback (RLHF) is a powerful method to further improve the quality of  trajectory datasets.

\textbf{Generic model with performance metrics:} Although replacing network experts is the ultimate goal of AIGN, training a generic AIGN model to cope with various network design scenarios is quite challenging. Apart from the convergence speed and average reward, metrics to measure the design performance of AIGN are still lacking. For example, the degree of task completion and the percentage of constraint satisfaction may be added to test the inference ability of AIGN.  Traditional solvers (e.g.,  CPLEX and Gurobi) can also be leveraged to verify and further optimize AIGN-designed solutions. Moreover, developing ChatGPT-based plug-and-play tools for AIGN to aid network design and management is a promising direction.

\textbf{Digital twins as interaction platforms:} Since directly deploying the AIGN to the real network environments can be expensive and very risky, combining AIGN with the digital twin is a potential research direction. The digital twin provides a real-time interaction platform between the physical infrastructure and the virtual counterpart, where we can continuously verify the effect of AIGN and synchronize the final designed solution with good performance to the physical entity. Moreover, with advanced analytics and modeling techniques, digital twin can also help reduce costs, improve efficiency, and accelerate the network development process, such as predictive maintenance, performance optimization, training, and testing. Additionally, unlike two-dimensional images, the exploration space of network problems grows exponentially with the number of users. Digital twin methods for visualizing network designs in high-dimensional space still remain to be investigated.

\section{Conclusions}

This article has merged the generative AI techniques and RL to enable AIGN, thereby earthing the potential of a data-driven paradigm for expert-free network optimization and promoting the value of intelligent decision-making in intent-driven network design. First, the AIGN framework with several appealing properties has been presented, and a novel diffusion model-based learning approach is proposed. Then,  a proof-of-concept prototype has been conducted to prove that AIGN can guide the network design of power allocation under wireless access scenarios. Finally, the challenges of the proposed AIGN are discussed. Since intelligence will play a defining role in the development of future networks, we hope AIGN can inspire follow-up research.

\bibliographystyle{IEEEtran}

\bibliography{IEEEabrv, reference.bib}

\begin{IEEEbiographynophoto}{Yudong Huang}
 received his B.S. degree from BUPT, China, in 2019. He is currently working toward his Ph.D. degree in BUPT. He is a visiting Ph.D. student with the School of Computer Science and Engineering, Nanyang Technological University, Singapore, from 2022. His current research interests include deterministic networks and network intelligence.
\end{IEEEbiographynophoto}

\vspace{1 mm}

\begin{IEEEbiographynophoto}{Minrui Xu} [S’23]
	received the B.S. degree from Sun Yat-Sen University, Guangzhou, China, in 2021. He is currently working toward the Ph.D. degree in the School of Computer Science and Engineering, Nanyang Technological University, Singapore. His research interests mainly focus on Metaverse, quantum information technologies, and deep reinforcement learning.
\end{IEEEbiographynophoto}

\vspace{1 mm}

\begin{IEEEbiographynophoto}{Xinyuan Zhang}
	 received her B.S. degree from BUPT, China, in 2019. She is currently working toward her Ph.D. degree in BUPT. She is a visiting Ph.D. student with the ISTD Pillar, Singapore University of Technology and Design, Singapore, from 2022. Her current research interests include satellite-terrestrial integrated networks and edge intelligence.
\end{IEEEbiographynophoto}

\vspace{1 mm}

\begin{IEEEbiographynophoto}{Dusit Niyato} [M'09, SM'15, F'17]
	is currently a professor in the School of Computer Science and Engineering, Nanyang Technological University, Singapore. He received the B.Eng. degree from KMITL, Thailand in 1999 and Ph.D. in electrical and computer engineering from the University of Manitoba, Canada in 2008. His current research interests include generative AI and network intelligence.
\end{IEEEbiographynophoto}

\vspace{1 mm}

 \begin{IEEEbiographynophoto}{Zehui Xiong} [M'20]
	 is an Assistant Professor at Singapore University of Technology and Design. Prior to that, he was a researcher with Alibaba-NTU Joint Research Institute, Singapore. He received the Ph.D. degree in Computer Science and Engineering at Nanyang Technological University, Singapore. His research interests include wireless communications and edge intelligence.
\end{IEEEbiographynophoto}

\vspace{1 mm}

 \begin{IEEEbiographynophoto}{Shuo Wang} 
	received his B.S. degree in communication engineering from Zhengzhou University, China, in 2013 and his Ph.D. degree from Beijing University of Posts and Telecommunications in 2018. He is currently an assistant professor at BUPT. His current research interests include  deterministic networks and network intelligence.
\end{IEEEbiographynophoto}

\vspace{1 mm}


 \begin{IEEEbiographynophoto}{Tao Huang} 
	 received the B.S degree from Nankai University, Tianjin, China, in 2002, the M.S. and Ph.D. degrees in communication and information system from the BUPT, China, in 2004 and 2007, respectively. He is currently a Professor with the BUPT. His current research interests include network architecture and network intelligence.
\end{IEEEbiographynophoto}

\end{document}